\documentclass[apj]{emulateapj}
\usepackage{natbib}
\usepackage{graphicx}
\def \lp{\>\> .}
\def \lc{\>\> ,}

\newcommand{\kms}{\mbox{\,km\,s$^{-1}$}}
\newcommand\cmv{\mbox{cm$^{-3}$}}
\newcommand\cmc{\mbox{cm$^{-2}$}}

\newcommand{\ta}{\mbox{$T_A^*$}}

\newcommand{\cplus}{C$^+$}

\def \CII{[C\,{\sc ii}]}

\def \Tex{$T_{\rm ex}$}

\def \lp{\>\> .}
\def \lc{\>\> ,}
\def \beq{\begin{equation}}
\def \eeq{\end{equation}}

\begin{document}
%

%%%%%%%%%

\title {Electron Excitation of High Dipole Moment Molecules Reexamined}
\author{Paul F. Goldsmith\altaffilmark{1} and Jens Kauffmann\altaffilmark{2}}

\altaffiltext{1}{Jet Propulsion Laboratory, California Institute of Technology, 4800 Oak Grove Drive, Pasadena CA, 
91109, USA; paul.f.goldsmith@jpl.nasa.gov}
\altaffiltext{2}{Max Planck Institut f\"{u}r Radioastronomie, Auf dem H\"{u}gel 69, D-53121 Bonn, Germany}

\begin{abstract}

Emission from high--dipole moment molecules such as HCN allows determination of the density in molecular clouds, and is often considered to trace the ``dense'' gas available for star formation.  We assess the importance of electron excitation in various environments.  The ratio of the rate coefficients for electrons and H$_2$ molecules, $\simeq$10$^5$ for HCN, yields the requirements for electron excitation to be of practical importance if $n({\rm H}_2) \leq\ 10^{5.5}$ \cmv\ and $X({\rm e}^-) \geq\ 10^{-5}$, where the numerical factors reflect critical values $n_{\rm{}c}({\rm H_2})$ and $X^*({\rm{}e}^-)$. This indicates that in regions where a large fraction of carbon is ionized, $X({\rm e}^-)$ will be  large enough to make electron excitation significant. The situation is in general similar for other ``high density tracers'', including HCO$^+$, CN, and CS. But there are significant differences in the critical electron fractional abundance, $X^*({\rm e}^-)$, defined by the value required for equal effect from collisions with H$_2$ and e$^-$.  Electron excitation is, for example, unimportant for CO and C$^+$.  Electron excitation may be responsible for the surprisingly large spatial extent of the emission from dense gas tracers in some molecular clouds (Pety et al.\ 2017; Kauffmann, Goldsmith et al.\ 2017). The  enhanced estimates for HCN abundances and HCN/CO and HCN/HCO$^+$ ratios observed in the nuclear regions of luminous galaxies may be in part a result of electron excitation of high dipole moment tracers. The importance of electron excitation will depend on detailed models of the chemistry, which may well be non--steady state and non--static.

\end{abstract}

\section{Introduction}
\label{intro}

The possible importance of excitation of the rotational levels of molecules by collisions with electrons, and consideration of the effect of such collisions on observed line ratios is not at all new \citep[e.g.][]{Dickinson77}.  However, relatively early  observations and modeling of molecular ions in dense clouds showed that in well--shielded regions, the fractional abundance of electrons is very low, 10$^{-7}$ -- 10$^{-8}$ \citep{Guelin77,Wootten79}.  Values of $X({\rm e}^-)$ in this range would make electron excitation insignificant, although the situation in clouds with lower extinction is dramatically different, with  $X({\rm e}^-)$ $\ge$ 10$^{-5}$ making electron collisions the dominant excitation mechanism for CS \citep{Drdla89} and for CN \citep{Black91}.  The early calculations of electron excitation rate coefficients \citep[e.g.][]{Dickinson75, Dickinson77} were forced to make a variety of approximations, but suggested that the excitation rates scale as the square of the molecule's permanent electric dipole moment.  Thus, electron excitation would be important for the widely--observed CO molecule only under very exceptional circumstances, but the question of the possible importance of electron excitation in varied regions of the interstellar medium has not been examined.  

In studies of star formation in other galaxies, emission from the high--dipole moment molecule HCN has been used as a measure of the ``dense'' gas in which star formation takes place \citep{Gao04}.  The question of the H$_2$ density that characterizes the regions responsible for this emission thus arises.  Since this emission is generally not spatially resolved, excitation by electrons could be contributing, especially in the outer regions of clouds subject to high  radiation fields.  \citet{Kauffmann17} have studied the density associated with HCN emission in the Orion molecular cloud, and find that a large fraction of the flux is produced in regions having $n({\rm{}H_2})\approx{}10^3~\cmv$, well below the range $\ge$\ 3$\times$10$^4$ \cmv\ assumed by \citet{Gao04}.    \citet{Pety17}  studied a variety of molecules including HCN in Orion B, and found that the spatial extent of their emission was not correlated with the density of H$_2$ required for collisional excitation.  One possible explanation is electron excitation in the outer regions of the cloud, making reexamination of the possible role of electron excitation appropriate
\footnote{We acknowledge appreciatively the suggestion by Simon Glover to anayze the possible role of electron excitation.}. 

In this paper, we review the recent rate calculations for HCN, $\rm{}HCO^+$, CS, and CN in \S \ref{rates}. Their influence on the excitation of molecules---with a particular focus on HCN---is summarized in \S \ref{sec:excitation}, which utilizes the three-level model developed in Appendix~\ref{app:simple} together with multilevel statistical equilibrium calculations. This section ends with a more general discussion that extends the argument to transitions of $\rm{}HCO^+$, CS, and CN (\S \ref{sec:extension}). Clouds of different types in different environments are examined using a PDR code to determine their electron density distribution in \S \ref{models}. In this section we also discuss the question of the abundance of molecules in the high--electron density regions including diffuse and translucent clouds, molecular cloud edges, and the central regions of active galaxies.  We summarize our conclusions in \S \ref{conc}.

\section{Collision Rate Coefficients}
\label{rates}  
We first discuss the HCN molecule as perhaps the premier example of a high--dipole moment molecule for which electron excitation can be relatively important.  Along with presenting their quantum collision rate coefficients for CS, \citet{Varambhia10} make a brief comparison of excitation by electrons and H$_2$, concluding that for $X({\rm e}^-)$ $\ge$ 10$^{-5}$, the latter should not be ignored.  To a reasonable approximation the collision cross sections for electron excitation will be dominated by long--range forces and scale as the square of the permanent electric dipole moment, $\mu_e$ \citep{Dickinson75, Dickinson77, Varambhia10}.  Thus the rates for the CS molecule having $\mu_e$ = 1.958 D \citep{Winnewisser68} or HF with $\mu_e$ = 1.827 D \citep{Muenter70} would be $\simeq$ 50\% of those of HCN having $\mu_e$ = 2.985 D \citep{Ebenstein84}.  An extremely polar molecule such as LiH, having $\mu_e$ = 5.88 D \citep{Buldakov04} would have electron collision rates almost a factor of 4 greater than those of HCN.  But overall, rates for high--dipole moment molecules are fairly well confined within about an order of magnitude.  The obvious outlier is CO, with $\mu_e$ = 0.11 D \citep{Goorvitch94}, thus having electron collision rate coefficients $\simeq$ 0.003 of those for high--dipole moment molecules.  In the following section we focus on HCN, given its observational importance. We also consider HCO$^+$, CS, and CN in \S \ref{HCO+dat}--\ref{CNdat}. 

\subsection{HCN Excitation by Electrons}
\label{HCNelecdat}
The calculation of electronic excitation of the lower rotational levels of HCN and isotopologues by \citet{Faure07} includes treatment of the hyperfine levels, and considered the HNC molecule and isotopologues as well.  Here, we do not consider the issue of the hyperfine populations, which although observable and informative in dark clouds with relatively narrow line widths, are not an issue for study of GMCs, especially large--scale imaging in the Milky Way and other galaxies.    

\citet{Faure07} present their results in the form of polynomial coefficients for the deexcitation ($J_{final} <  J_{initial}$) rate coefficients as a function of the kinetic temperature.  We have calculated rates for a number of temperatures and give the results for some of the lowest rotational transitions in Table \ref{drates}.  We include here as well the analogous results for HCO$^+$, CS, and CN, which are discussed in \S\ref{HCO+dat}--\ref{CNdat}. We see that the temperature dependence of the deexcitation rate coefficients is quite weak, and that in common with previous analyses, $| \Delta J |$ = 1 (dipole--like) transitions are strongly favored for electron excitation of neutrals, but less strongly so for electron excitation of ions.  For any transition, the collision rate is the product of the collision rate coefficient and the density of collision partners (e.g. electrons or H$_2$ molecules); $C$(s$^{-1}$) = $R$(cm$^3$s$^{-1}$)$n$(e$^-$ or H$_2$; cm$^{-3}$).  The full set of deexcitation rate coefficients is available on the LAMBDA website (\citet{Schoier05}; \path{http://home.strw.leidenuniv.nl/~moldata/}).

%%%%%%%%%%%%%%%%%%%%%%%%%%%%%%%%%%%%%%%
% Table 1
%%%%%%%%%%%%%%%%%%
\begin{deluxetable}{lclllll}[ht!]
\tablewidth{0pt}
\tablecaption{Electron Deexcitation Rate Coefficients for Lower Rotational Transitions of HCN\tablenotemark{1}, HCO$^+$\tablenotemark{2}, and CS\tablenotemark{3} (Units are $10^{-6}$ cm$^{3}$ s$^{-1}$) \label{drates}} 
\tablehead{  
\colhead{Molecule}&\colhead{Transition} &\multicolumn{5}{c}{Kinetic Temperature (K)}\\
& \colhead{$J_u$ -- $J_l$}  & \colhead{10} & \colhead{20} & \colhead{40} & \colhead{80} & \colhead{100}
}
% 10 20 40 80 100
\startdata
HCN 			& 1 -- 0 & 3.9 & 3.7 & 3.5 & 3.2 & 3.1\\
        			& 2 -- 1 & 3.7 & 3.6 & 3.4 & 3.1 & 3.0\\
	   			& 2 -- 0 & 0.089 & 0.076 & 0.064 & 0.054 & 0.051\\
HCO$^+$ 	& 1 -- 0 &13.5&9.5 & 6.6 & 6.0 &3.8  \\
	   			& 2 -- 1 &15.3&10.8& 7.8  &5.7   &5.0 \\  
	   			& 2 -- 0 &1.5  & 1.0 & 0.75& 0.53&0.46\\
CS    			& 1 -- 0 &1.8  & 1.7 & 1.6 & 1.6 & 1.5\\
        			& 2 -- 1 &1.9  & 1.8  & 1.7 & 1.6 & 1.6\\
	   			& 2 -- 0 &0.044&0.038&0.031&0.024&0.023
	   
\enddata

\tablenotetext{1}{see \S \ref{HCNelecdat}}
\tablenotetext{2}{see \S \ref{HCO+dat}}
\tablenotetext{3}{see \S \ref{CSdat}}

\end{deluxetable}
%%%%%%%%%%%%%%%%%%%%%%%%%%%%%%%%%%%%%%%%%%%%%%%%%

\subsection{HCN Excitation by H and H$_2$}

The calculation of rate coefficients for collisions between HCN and H$_2$ molecules started with \citet{Green74}, who considered HCN as having only rotational levels and included He as the collision partner, representing H$_2$ in its ground para--H$_2$ (I = 0) state with antiparallel nuclear spins.  Interest in the non--LTE ratio of HCN hyperfine components led \citet{Monteiro86} to include the hyperfine levels separately.  They found that the individual excitation rate coefficients summed together to give total rotational excitation rate coefficients very similar to those found by \citet{Green74}. \citet{Sarrasin10} again considered He as the collision partner, while focusing on differences between the excitation rates for HCN and HNC. 

\citet{Dumouchel10} employed a new potential energy surface (PES), while still considering the collision partner to be He.  The rate coefficients are not very different from those found previously, but they do confirm the difference between HCN and HNC.  \citet{BenAbdallah12} treat the colliding H$_2$ molecule as having internal structure, but average over H$_2$ orientations, considering effectively only molecular hydrogen in the $j$ = 0 level (we employ lower case $j$ to denote the rotational level of H$_2$ in order to avoid confusion with the rotational level of HCN).

\citet{Vera14} have recently  calculated collisions between HCN and H$_2$ molecules, considering for the first time the latter in individual rotational states.  They find that there is a significant difference between collisions with the H$_2$ in the $j$ = 0 level, compared to being in higher rotational levels.  The deexcitation rate coefficients for the lower levels of HCN for H$_2$($j$ $\ge$ 1) are quite similar, and are 3 to 9 times greater than those for H$_2$($j$ = 0), rather than, for example, there being a systematic difference for ortho-- and para--H$_2$ rates.  The HCN deexcitation rates for H$_2$($j$ = 0) are generally similar in magnitude to those of \citet{Dumouchel10} and \citet{BenAbdallah12}; the deexcitation rates for $\Delta J$  = --1 and --2 transitions are comparable, in contrast to those for H$_2$($j \ge $ 1), which show a significant propensity for $\Delta J$ = --2.  The numerical results for many of these calculations (often not given in the published articles) are available on the website \path{http://basecol.obspm.fr}.

The major difference between the rates for H$_2$($j$ = 0)  and H$_2$($j$ $\ge$ 1) adds a significant complication to the analysis of HCN excitation since it implies a dependence on the H$_2$ ortho to para ratio, which is itself poorly--known, and likely varies considerably as a function of environment \citep[e.g.][]{Neufeld06, Maret09}.  If the H$_2$ ortho to para ratio is close to equilibrium at the local kinetic temperature, then in all but the most excited regions of molecular clouds, only the collision rate coefficients with H$_2$($j$ = 0) will be significant.

%%%%%%%%%%%%%%%%%%%%%%%%%%%%%%%%%%%%%%
\subsection{HCO$^+$}
\label{HCO+dat}
\citet{Faure01} considered electron excitation of the HCO$^+$ ion.  This work provided coefficients for evaluating the rates only for the three lowest rotational states, but the rates themselves are comparable to those of HCN, and thus are reasonably consistent with a scaling following $\mu_e^2$.  This calculation was supplemented by one including more levels with improved accuracy at low temperatures described by \citet{Fuente08} and kindly provided to us by A. Faure.  The newer deexcitation rate coefficients for low--J transitions are a factor $\simeq{}3$ larger at 10 K, but the difference drops rapidly for higher kinetic temperatures and is only 10 to 20\% for $T_k$ = 100 K.  We include deexcitation rate coefficients in Table \ref{drates}.  Collision rates for HCO$^+$ excitation by H$_2$ have been calculated by \citet{Flower99}.  The deexcitation rate coefficients for collisions are a factor of 3 to 25 times larger than those for HCN.  The HCO$^+$ $\Delta J$ = --1 rate coefficients are larger than those for $\Delta J$ = --2 collisions, unlike the case for HCN, for which the inverse relationship holds.

%%%%%%%%%%%%%%%%%%%%%%%%%%%%%%%%%%%%%%%%%%
\subsection{CS}
\label{CSdat}
Electron collisions for CS have been analyzed by \citet{Varambhia10}, who found deexcitation rate coefficients a factor of 2 smaller than those for HCN, and with an exceptionally strong propensity rule favoring $\Delta J$ = --1 collisions.   A selection of the lowest transitions are included in Table \ref{drates}.   The various calculations for electron excitation of the lower transitions of CS by various methods vary by less than 30\%, and less for temperatures $\geq$ 50 K \citep{Varambhia10}, a typical accuracy that is probably characteristic of the electron collision rate coefficients for other molecules.   Deexcitation rate coefficients for collisions with H$_2$ are from \citet{Lique06}, but these results do not differ appreciably from those of \citealt{Turner92}) and are $\simeq$ 3$\times$10$^{-11}$ cm$^3$s$^{-1}$, comparable to those for HCN for $\Delta J$ = --2, but a factor 2 to 3 larger for $\Delta J$ = --1. 

%%%%%%%%%%%%%%%%%%%%%%%%%%%%%%%%%%%%%%%
\subsection{CN}
\label{CNdat}
The spin--rotation coupling for CN complicates the energy level structure and makes accurate comparisons with simple rotors difficult.  Excitation rate coefficients for collisions with electrons were calculated by \citet{Allison71} for the lowest few levels, and extended by \citet{Black91}.  From the Lambda Leiden Molecular Data Base (\path{http://home.strw.leidenuniv.nl/~moldata/datafiles/cn.dat}) we find a characteristic rate coefficient at 20 K for $N$ = 1--0 equal to 5.7$\times$10$^{-7}$ cm$^3$s$^{-1}$.  
%Rate coefficients for collisions with H$_2$ have been obtained by \citet{Lique10}, who scaled values calculated for collisions with He by a factor 1.37, finding  the deexcitation rate coefficient for $N$ = 1--0 equal to 7.2$\times$10$^{-12}$ cm$^3$s$^{-1}$.  

%%%%%%%%%%%%%%%%%%%%%%%%%%%%%%%%%%%%%%%%%
%%%%%%%%%%%%%%%%%%%%%%%%%%%%%%%%%%%%%%%%%
\section{Excitation by Electrons and H$_2$ Molecules}
\label{sec:excitation}
In this section we discuss how molecules are excited in collisions with electrons and $\rm{}H_2$ molecules. Throughout this section we often use HCN as a reference case, given the high astrophysical importance of the molecule and its high sensitivity to collisions with electrons.

\subsection{Critical Electron Fractional Abundance\label{sec:critical-fractional-abundance}}
The critical density of a molecule is often used to indicate how a molecular species depends on the environmental density. Appendix~\ref{app:simple} provides a discussion of such trends. For our discussion it is important to realize that we are dealing with two critical densities per molecule: one critical value that describes the excitation with electrons, $n_{\rm{}c}({\rm{}e}^-)$, and one that describes the excitation in collisions with $\rm{}H_2$ molecules, $n_{\rm{}c}({\rm{}H_2})$. These densities can then be used to gauge the relative importance of electron excitation for the excitation of a given molecule. In the context of the simplified 3--level model described in Appendix~\ref{app:simple}, the critical fractional abundance of electrons required to have the electron collision rate be equal to the H$_2$ collision rate is
\beq
X^*({\rm e}^-) =  \frac{R^e_{10}({\rm H}_2)}{R^e_{10}({\rm e}^-)} =   \frac{n_{\rm c}({\rm e}^-)}{n_{\rm c}({\rm H}_2)} \lp
\eeq
 
Table \ref{critical} gives the critical densities and critical fractional abundance of electrons for the $J$ = 1--0 transitions of HCN, HCO$^+$, CS, and CN.  The entries for CN are representative values for the $N$ = 1--0 transitions (near 113 GHz) and these must be regarded as somewhat more uncertain due to more complex molecular structure and less detailed calculations, as discussed above.  HCN has relatively small rate coefficients for collisions with H$_2$ and consequently large $n_c({\rm H}_2)$, while only modestly smaller rate coefficients for collisions with electrons.  The result is that the critical electron fraction for HCN is lower than for the other high--dipole moment species considered.  CN follows, with CS having a somewhat higher value of $X^*({\rm e}^-$).  HCO$^+$ has a rather significantly higher value yet, making this species less likely to be impacted by electron excitation than the others.
%%%%%%%%%%%%%%%%%%%%%%%%%%%%%%%%%%%%%%%
%%  Now Table 2
%%%%%%%%%%%%%%%%%%%%%%%%%%%%%%%%%%%%%%%
\begin{deluxetable}{lccc}[ht!]
\tablewidth{0pt}
\tablecaption{Critical Densities and Critical Electron Fractional abundances for the J = 1--0 Transitions\tablenotemark{1} of Different Species at 20 K \label{critical}
} 
\tablehead{  
\colhead{Molecule} &
\colhead{$n_c(  {\rm e}^-$)} & 
\colhead{$n_c(  {\rm H}_2$)} & 
\colhead{$X^*({\rm e}^-$)}\\
\colhead{}&
\colhead{cm$^{-3}$} & 
\colhead{cm$^{-3}$} & 
\colhead{}
}
\startdata
HCN 			&6.5   &6.5$\times$10$^5$ &1.0$\times$10$^{-5}$ \\
HCO$^+$	&4.5	&1.2$\times$10$^5$ &3.8$\times$10$^{-5}$\\
CS				&1.8	&2.3$\times$10$^4$ &7.9$\times$10$^{-5}$\\
CN				&21	&1.7$\times$10$^6$ &1.3$\times$10$^{-5}$
\enddata
\tablenotetext{1}{The entries for CN are characteristic values for the $N$ = 1--0 transitions, and are somewhat approximate}
\end{deluxetable}

\subsection{Emission and Excitation Temperature\label{sec:emission-tex}}
A complementary approach is to consider how the integrated intensity of different species is affected by electron excitation.  For optically thin emission, the integrated antenna temperature is just proportional to the upper level column density which for a uniform cloud this is proportional to the upper level density.  In the low density limit with no background radiation, from equations \ref{lowdenratio} and \ref{geneffective} we can write (for excitation by any combination of collision partners) 
\beq
\label{Tdv_Ct}
\int T_{\rm a} dv \propto N_1A_{10} = N({\rm HCN})C^{\rm t}_{01} \lc
\eeq
where $N_1$ is the column density of HCN in the $J$=1 state,  $N$(HCN) is the total column density of the molecule, and 
\beq
\label{Ct_def}
C^{\rm t}_{01} = C^{\rm e}_{01}({\rm e}^-)  + C^{\rm e}_{01}({\rm H}_2)
\eeq
is the total collisional excitation rate from $J$ = 0 to $J$ = 1.  The total deexcitation rate is determined from equation\ref{Ct_def} through detailed balance.

Using the relationship between the collision rates and collision rate coefficients for the electrons and H$_2$ molecules and the critical densities for each species (from equations \ref{elec_nc} and \ref{H2_nc}), we can express the integrated intensity as 
\beq
\label{Tdv_crit}
\int T_{\rm a} dv \propto N({\rm HCN}) A_{10} 
 \left( \frac{n({\rm{}e}^-)}{n_{\rm{}c}({\rm{}e}^-)} + \frac{n({\rm{}H}_2)}{n_{\rm{}c}({\rm{}H}_2)} \right) \, .
\eeq
%
% took out factor \frac{g_1}{g_0} \exp(-T^*_{10}/T_k)  as it just clutters things up
%
A measure of the degree of excitation of the $J$ = 1--0 transition is its excitation temperature, which is defined by the ratio of molecules per statistical weight in the upper and lower levels.  With $T^*_{10}$ = $\Delta E/k_B$ we can write
\beq
T_{{\rm ex}_{10}} = 
  \frac{
    T^*_{10}
  }{
    \ln \left( \frac{N_0g_1}{N_1g_0} \right)
  } = \frac{
    4.25~{\rm K}
  }{
    \ln \left( \frac{3n_0}{n_1} \right)
  } = \frac{
    4.25~{\rm K}
  }{
    \ln \left( \frac{3A_{10}}{C^t_{01}} \right)
  } \lc
\eeq
where $g_0$ = 1 and $g_1$ = 3.

\begin{figure}[ht!]
\begin{center}
\includegraphics[angle=0, width=\linewidth]{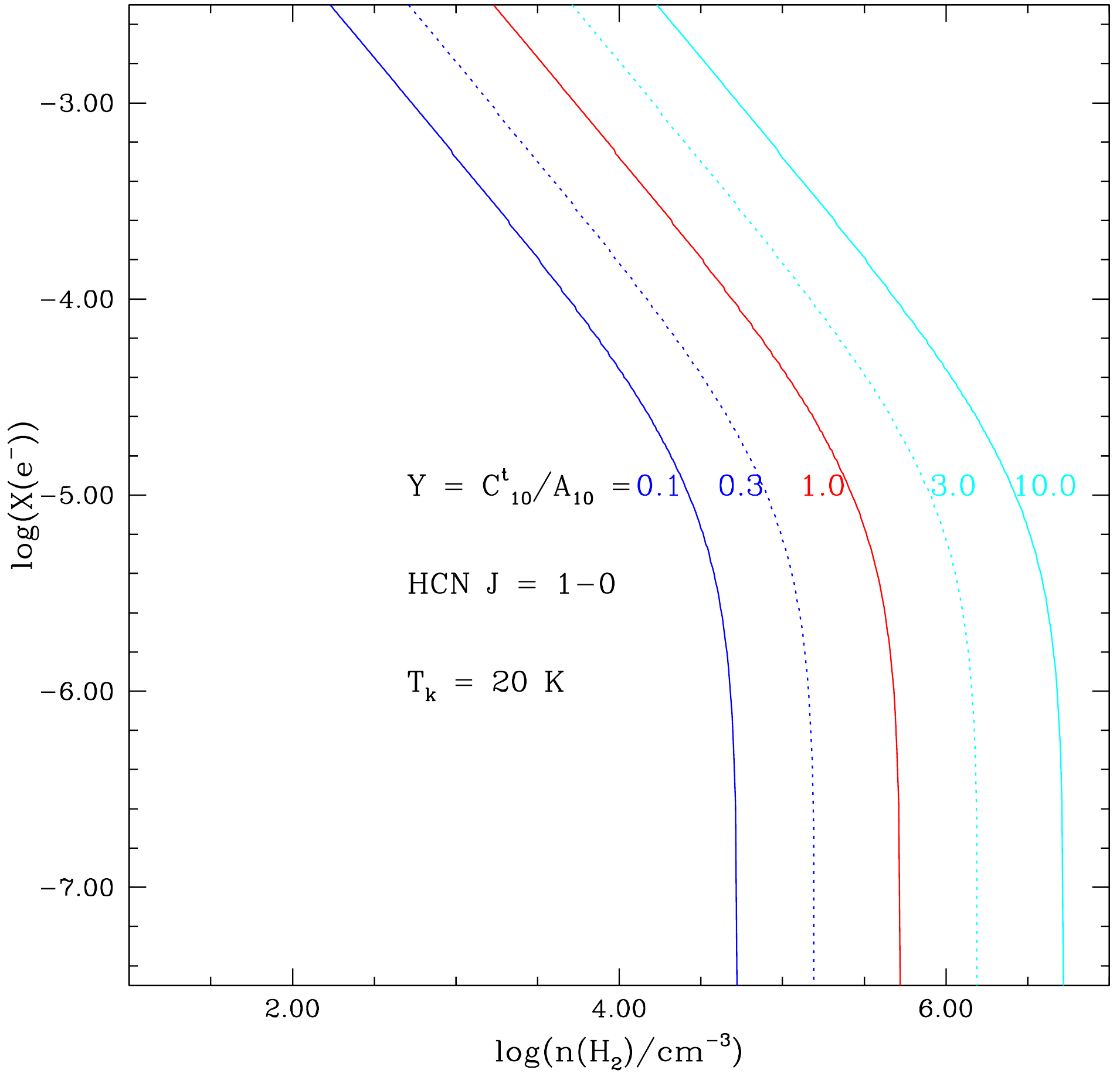}
\caption{Curves showing the value of the electron fractional abundance as function of the H$_2$ density required to achieve the indicated values of the the thermalization parameter $Y = C^t_{10}/A_{10}$.  In the area where the curves are vertical, electron excitation is unimportant, and in the area where the curves are diagonal, the electrons are the major contributor to the total collision rate. }
\label{HCN_nX}
\end{center}
\end{figure}

The importance of electron collisions for the excitation of HCN (or other high--dipole moment molecules) does depend on the rate of neutral particle excitation that is present, as shown in Figure \ref{HCN_nX}.  
%Here, we define a total deexcitation rate $C^t_{10} = C^e_{10}({\rm H}_2) + C^e_{10}({\rm e}^-)$
The thermalization parameter $Y = C^t_{10}/A_{10}$ gives the total deexcitation rate relative to the spontaneous decay rate. Each of the curves in Figure \ref{HCN_nX} is for a given value of $Y$, with small values of $Y$ indicating subthermal excitation and $Y \gg 1$ indicating thermalization.

In the area where the curves are essentially vertical, the H$_2$ density is sufficient to provide the specified value of $Y$ and the fractional electron abundance is sufficiently small that electrons are unimportant collision partners.  In the area in which the curves run diagonally, the value of $Y$ increases linearly as a function of $X({\rm e}^-)$ and $n({\rm H}_2)$, indicating that electrons are the dominant collision partners.  

In order that electron collisions be of practical importance, we must satisfy two conditions.  First, the H$_2$ density must be insufficient to thermalize the excitation temperature, meaning that $n({\rm H}_2) \leq\ n_{\rm{}c}({\rm{}H_2})$.  For HCN J = 1--0 this implies $n({\rm H}_2) \leq\ 10^{5.5}$.  Second, $X({\rm e}^-)$ must be sufficiently large that electrons are the dominant collision partner. This means that we must be in or near to the area of the diagonal curves, which in combination with requirement 1 for that molecular transition we take as defined by $X({\rm e}^-) \geq\ X^*({\rm e}^-)$. For HCN J = 1--0 this means $X({\rm e}^-) \geq\ 10^{-5}$. 
%%%%%%%%%%%%%%%%%%%%%%%%%%%%%%%%%%%%%
%% FIGURE 1
%%%%%%%%%%%%
%%%%%%%%%%%%%%%%%%%%%%%%%%%%%%%%%%%%%%%%%
%%%%%%%%%%%%%%%%%%%%%%%%%%%%%%%%%%%%%%%%%
\subsection{Multilevel Results for HCN}

\begin{figure}[h!]
\begin{center}
\includegraphics[angle=0, width=\linewidth]{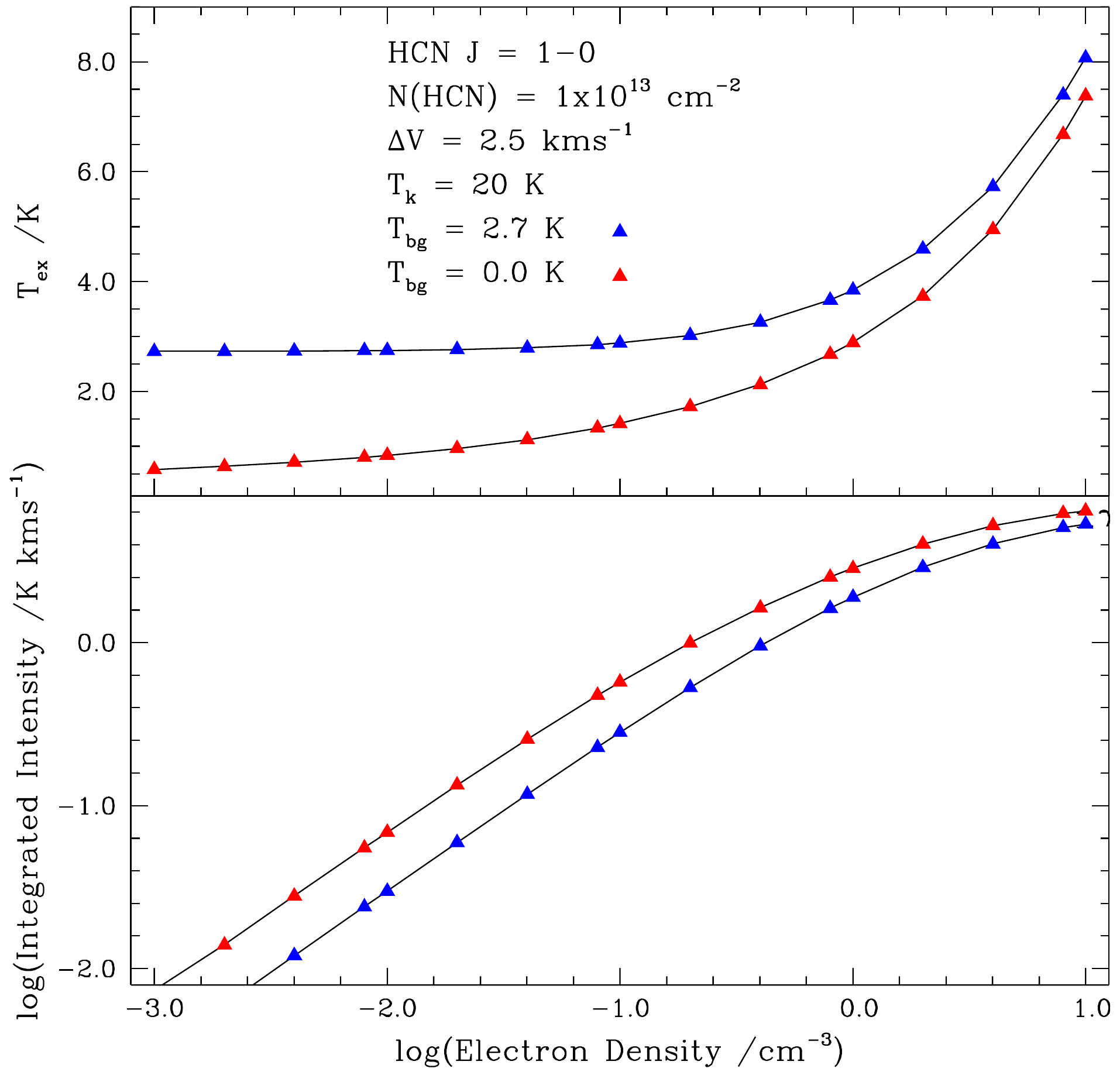}
\caption{Purely electron excitation of the HCN $J$ = 1--0 transition.  The kinetic temperature is 20 K, the HCN column density is 10$^{13}$ \cmc, and the line width is 2.5 \kms.  Upper panel:  excitation temperature; Lower panel:  Integrated line intensity.  The points in blue are for a background temperature equal to 2.7 K and those in red for no background.}
\label{elec_0.0}
\end{center}
\end{figure}

In Figure \ref{elec_0.0} we show the results for purely electron excitation of the HCN $J$ = 1--0 transition from a 10 level calculation using RADEX \citep{Vandertak07} for the indicated conditions.  In the upper panel we show the excitation temperature as a function of the electron density for the cases of a background temperature equal to 2.7 K (blue symbols) and  equal to 0.0 K (red triangles).  Given that the equivalent temperature difference between the upper and lower levels is 4.25 K, the higher background temperature corresponds to a significant excitation rate, and \Tex\ rises above the background only for an electron density of a few tenths \cmv.  In the case of no background, however, there is no ``competition'' for the collisional excitation, and \Tex\ increases monotonically starting from the lowest values of the electron density.  

%%%%%%%%%%%%%%%%%%%%%%%%%%%%%%%%%%%%%
% FIGURE 2
%
%%%%%%%%%%%%%%%%%%%%%%%%%%%%%%%%%%%%%%%

In the lower panel we show the integrated intensity of the $J$ = 1--0 line.  With or without background, the emission increases linearly with collision rate as expected, as long as $n({\rm e}^-)$ is well below $n_c({\rm e}^-)$ = 6.5 \cmv\ (Equation \ref{elec_nc}).  The nonzero background temperature reduces the integrated intensity by a constant factor due to the reduced population in the $J$ = 0 level available for excitation and emission of photons \citep{Linke77}.

\begin{figure*}
\begin{center}
\centerline{\includegraphics[angle=0, width= 0.7\textwidth]{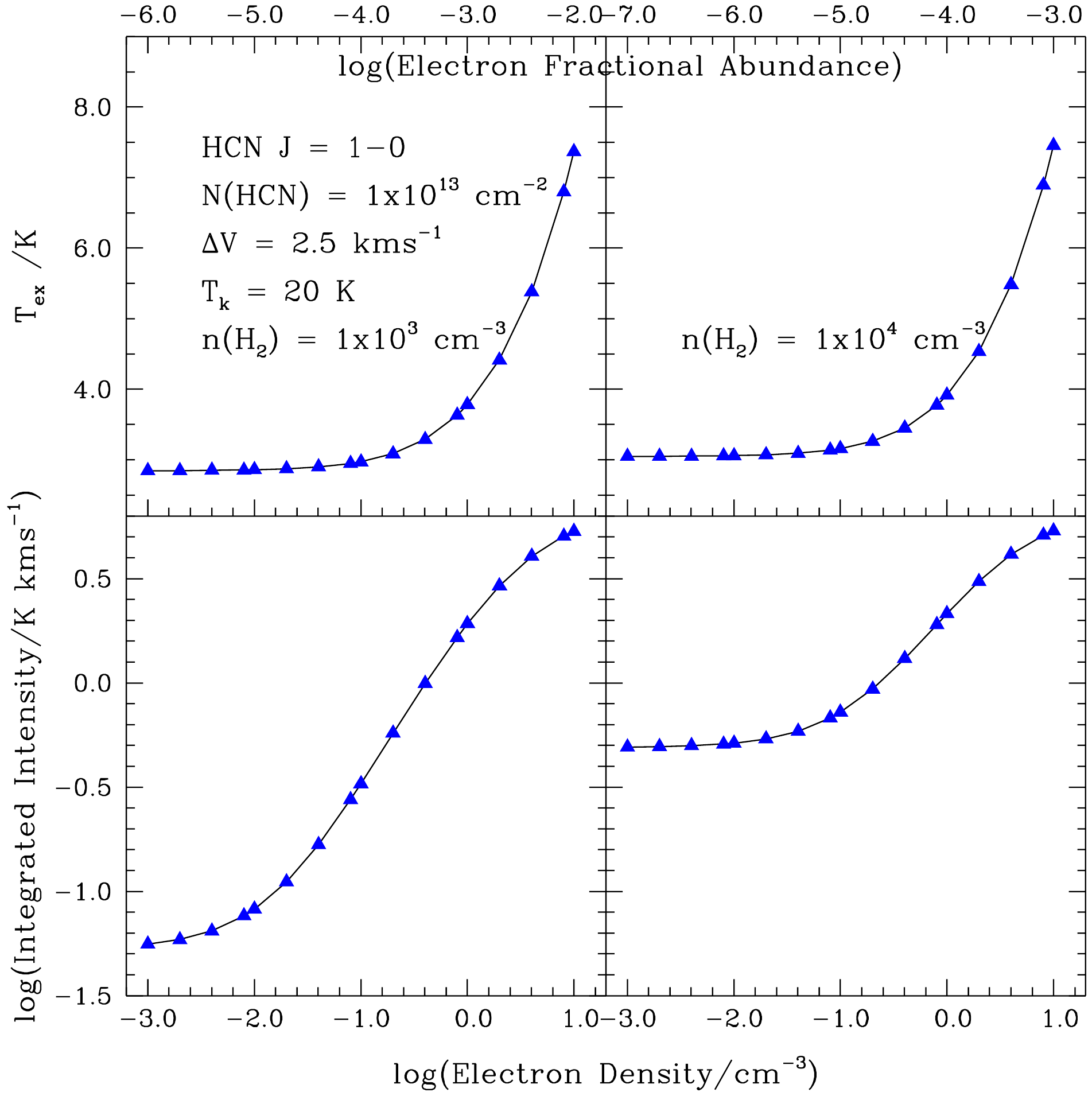}}
\caption{Effect of electron excitation on the $J$ = 1--0 transition of HCN.  The kinetic temperature is 20 K, the HCN column density is 10$^{13}$ \cmc, the line width is 2.5 \kms, and the background temperature is 2.7 K.  The left hand panels are for H$_2$ density equal to 10$^3$ \cmv\ and the right hand panels for H$_2$ density equal to 10$^4$ \cmv.  In each column the upper panel shows the  excitation temperature and the lower panel the integrated line intensity.}
\label{elec_1e31e4}
\end{center}
\end{figure*}

In Figure \ref{elec_1e31e4} we present the results of a multilevel calculation for the conditions indicated.  
%%%%%%%%%%%%%%%%%%%%%%%%%%%%%%%%%%%%%
%FIGURE 3  - now combined for two densities
%%
%%%%%%%%%%%%%%%%%%%%%%%%%%%%%%%%%%%%%%%
The H$_2$ excitation provides a certain level of excitation and emission; this is seen most clearly by comparing the left and right lower panels showing the integrated intensity.   The excitation temperature for low electron densities is dominated by the background radiation and for both of the H$_2$ densities considered, the collisional excitation rate is not sufficient to increase $T_{ex}$ significantly above  2.7 K.  In the presence of the background, the integrated emission is a more sensitive reflection of electron excitation than the excitation temperature.  In agreement with the previous approximate analysis, the effect of the electron collisions becomes significant when the electron fractional abundance $X({\rm e}^-$) reaches 10$^{-5}$ (for $n$(H$_2$) = 10$^3$ or 10$^4$ \cmv), at which point the emission has increased by 50\%.  Electron excitation is dominant for $X({\rm e}^-$) = 10$^{-4}$, with the integrated intensity increasing by a factor of 5.5 for $n$(H$_2$) = 10$^3$ \cmv, and by a factor of 4.5 for $n$(H$_2$) = 10$^4$ \cmv.
For an H$_2$ density equal to 10$^4$ \cmv\ the H$_2$ excitation is significantly greater due to the order of magnitude greater density, but the electron density required to reach a level of emission significantly greater than that produced by the H$_2$ alone (e.g. $\int$\ta$dv$ = 3 K \kms) is independent of the H$_2$ density.

\subsection{Extension to HCO$^+$, CS, and CN}
\label{sec:extension}
\begin{figure}[ht!]
\begin{center}
\includegraphics[angle=0, width=\linewidth]{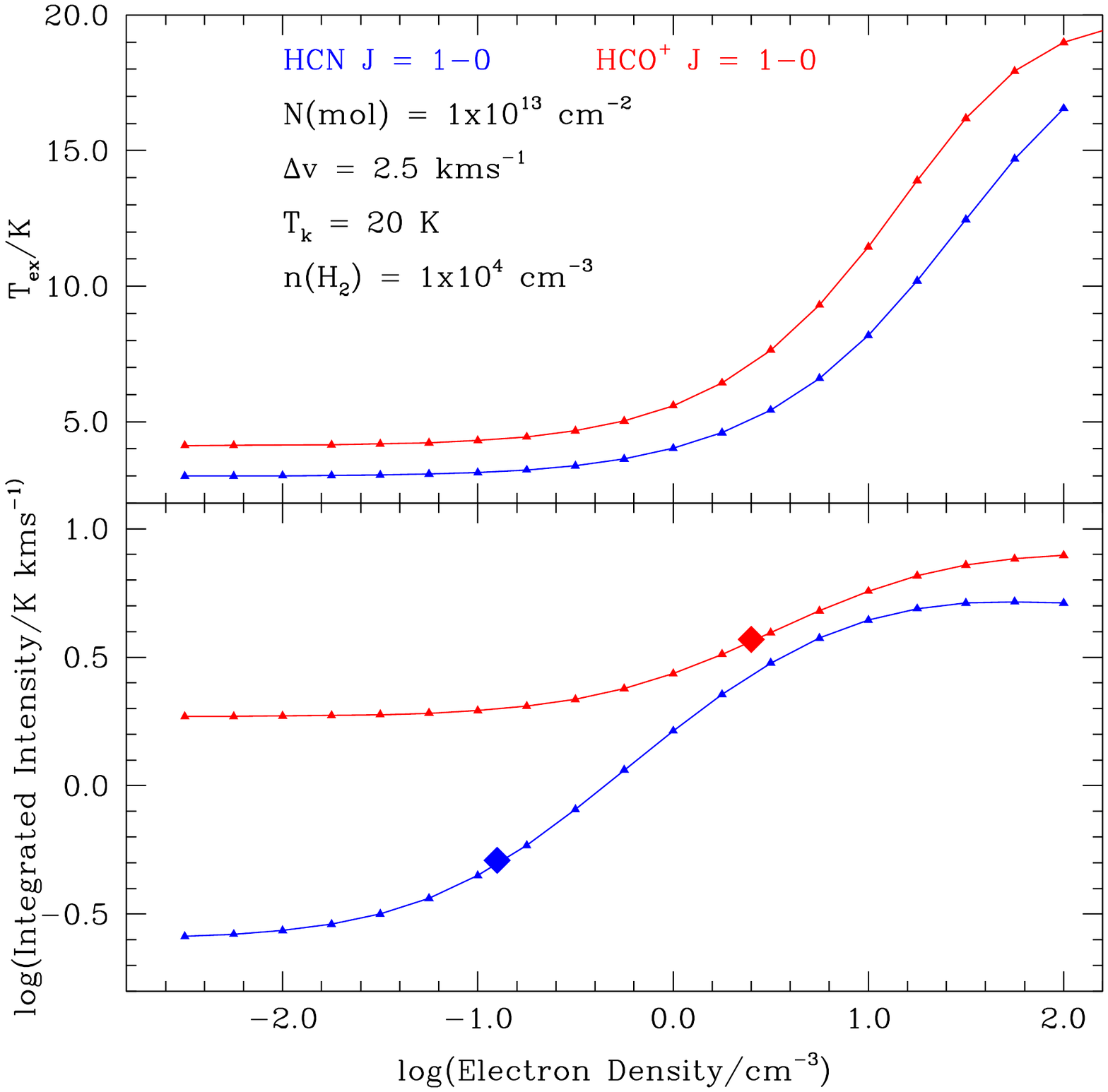}
\caption{Comparison of HCN and HCO$^+$ excitation by H$_2$ molecules and by electrons. The excitation temperature (upper panel) and integrated intensity (lower panel) of the J = 1--0 transition of HCN (blue) and HCO$^+$ (red) are shown as a function of the electron density for a H$_2$ density of 10$^4$ \cmv.  The diamonds indicate the electron density for which the integrated intensity is doubled as a result of increased collisional excitation by electrons.}
\label{HCN_HCO+}
\end{center}
\end{figure}

It is more difficult for electron excitation to play a role for HCO$^+$ than for HCN, since a much higher fractional electron abundance is required in order that the electron rate be comparable to or exceed that for H$_2$.  This is shown by the offset in the electron fractional abundances of the diamond symbols in Figure \ref{HCN_HCO+} which show the value of X(e$^-$) required to double the intensity of the J = 1--0 transition relative to that produced by collisions with H$_2$.  A fractional abundance of electrons $\simeq$ 15 times greater is required for HCO$^+$ relative to that for HCN, largely due to the far larger H$_2$ deexcitation rate coefficients for HCO$^+$ more than outweighing its only somewhat larger e$^-$ deexcitation rate coefficients.
%%%%%%%%%%%%%%%%%%%%%%%%%%%%%%%%%%%%%%%%%%
%%%  FIGURE 4
%%%%%%%%%%%%%%%%
%%%%%%%%%%%%%%%%%%%%%%%%%%%%%%%%%%%%%%%%%%

\begin{figure}[ht!]
\begin{center}
\includegraphics[angle=0, width=\linewidth]{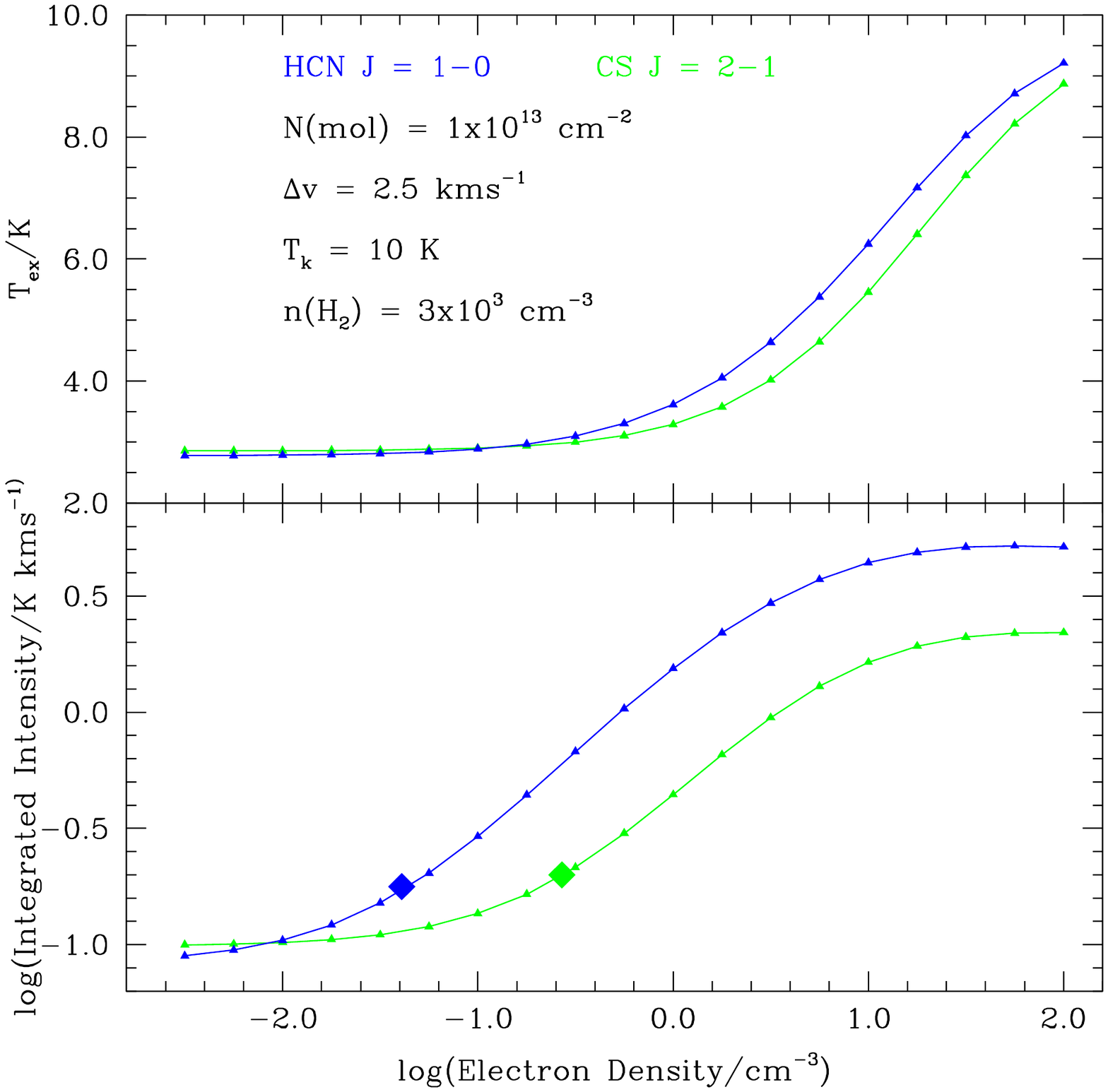}
\caption{Comparison of HCN and CS excitation by H$_2$ molecules and by electrons. The excitation temperature (upper panel) and integrated intensity (lower panel) of the J = 1--0 transition of HCN (blue) and the J = 2--1 transition of CS (green) are shown as a function of the electron density for a H$_2$ density of 3$\times$10$^3$ \cmv\ and a kinetic temperature of 10 K.  The diamonds indicate the electron density for which the integrated intensity is doubled as a result of increased collisional excitation by electrons. }
\label{HCN_CS}
\end{center}
\end{figure}

Figure \ref{HCN_CS} compares HCN and CS excitation as a function of electron density for a H$_2$ density of  3$\times$10$^3$; this lower density is appropriate to ensure subthermal excitation.  We consider the J = 2--1 transition of CS and the J = 1--0 transition of HCN in order to ensure comparable spontaneous decay rates.  We see that an electron density $\simeq$ 7 times greater for CS than for HCN is required to produce a factor of 2 enhancement in the integrated intensity.  

%%%%%%%%%%%%%%%%%%%%%%%%%%%%%%%%%%%%%%%%
%%   FIGURE 5
%%%%%%%%%%%%%%%
%%%%%%%%%%%%%%%%%%%%%%%%%%%%%%%%%%%%%%%%%

The conclusion from comparison of CS and HCO$^+$ with HCN is that for the latter molecule, a significantly lower fractional abundance of electrons can result in doubling the integrated intensity of the emission.  Thus, if electron excitation is significant, we might expect enhanced HCN to  HCO$^+$ ratio, more extended HCN emission, or both, and the same, though to a lesser degree, relative to CS.  However these conclusions are highly dependent on the chemistry that is determining the abundances of these species.

\section{Cloud Models and and the Electron Abundance}
\label{models}
\subsection{ Diffuse and Translucent Clouds}
\label{diffuse_trans}
\begin{figure}[h!]
\begin{center}
\includegraphics[angle=0, width=\linewidth]{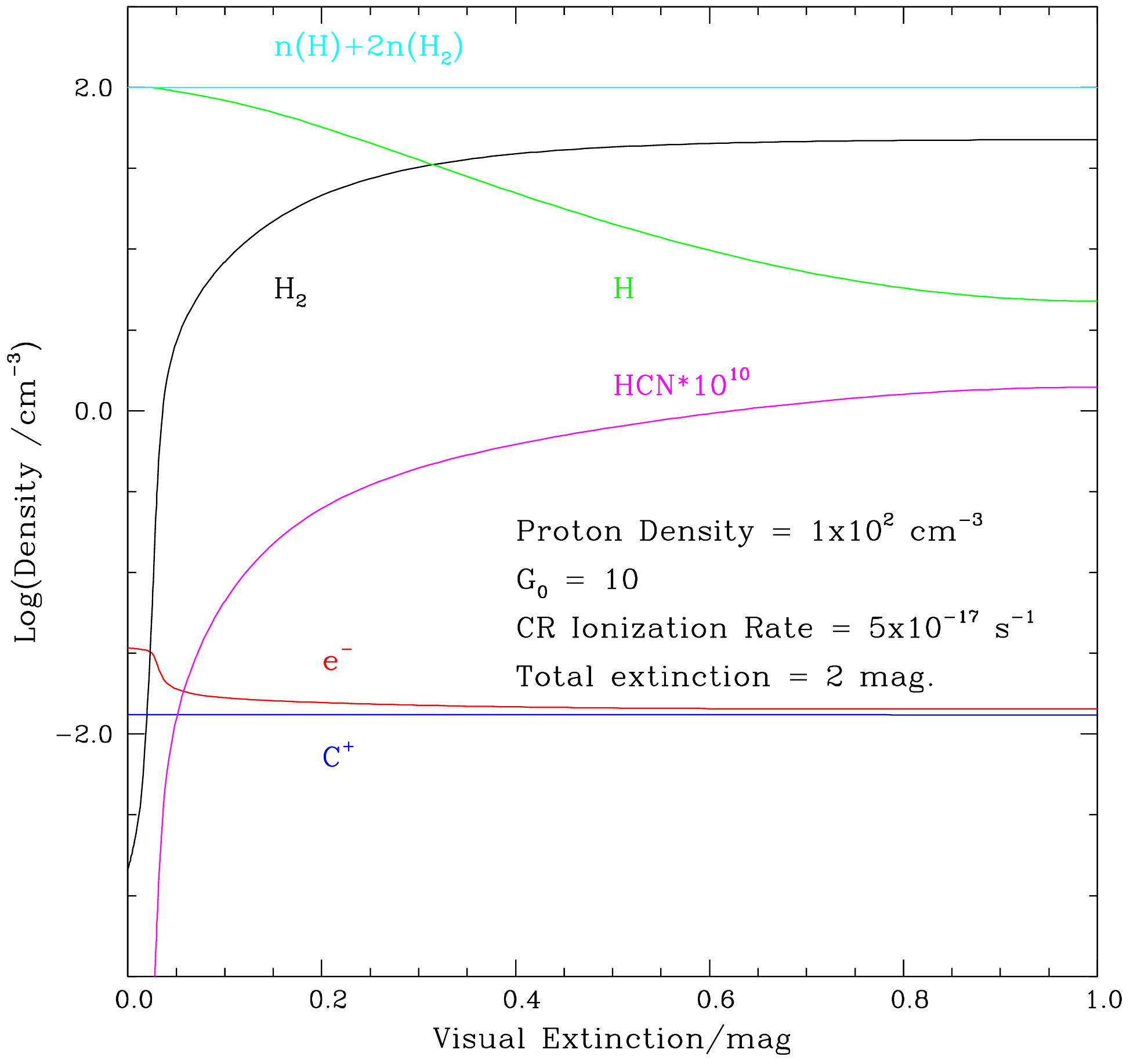}
\caption{One half of a diffuse cloud modeled as a uniform density slab having total extinction = 2.0 mag., irradiated on both sides, with parameters indicated.  The electrons present come primarily from ionized carbon, except in the outer 0.05 mag. of the cloud.}
\label{diffuse}
\end{center}
\end{figure}

Diffuse and translucent clouds have modest total extinction ($A_{\rm v}$ $\leq$ 2 mag.) and densities typically 50--100 \cmv.  In consequence, carbon is largely ionized and the electron fractional abundance is on the order of 10$^{-4}$.  Thus, as mentioned in \S \ref{intro}, electron excitation of high--dipole moment molecules will be very significant.  We have used the Meudon PDR code \citep{Lepetit06} to calculate the thermal and chemical structure of this cloud and show the results in Figure \ref{diffuse}.  Hydrogen is largely molecular except in the outer 0.25 mag. of the cloud and the electron abundance of 0.01 \cmv\ results in $n$(e$^-$)/$n$(H$_2$) = 2$\times$10$^{-4}$ throughout most of the cloud\footnote{All of the Meudon PDR code results presented in this paper assume a carbon to hydrogen ratio equal to 1.3$\times$10$^{-4}$.  This is somewhat lower than the value 1.6$\times$10$^{-4}$ obtained for four sources by \citet{Sofia04} using UV observations of \cplus\ absorption, and the value 1.4$\times$10$^{-4}$ adopted for analysis of the \CII\ 158 $\mu$m fine structure line by \citet{Gerin15}.  Measurements of carbon and oxygen abundances in ionized regions compiled by \citet{Esteban13} suggest a significant gradient in the [C]/[H] ratio, which they determine to be 6.3$\times$10$^{-4}$ at a galactocentric distance of 6 kpc and 2.5$\times$10$^{-4}$ at 10.5 kpc.  A higher carbon abundance translates to higher electron abundance where carbon is ionized, so that we may be underestimating the importance of electron excitation, but by an amount that likely depends on environment and location.}.
% 0.01/50 = 2e-4 
Excitation of high--dipole moment molecules will thus predominantly be the result of collisions with electrons. However, as seen in the Figure, the density of HCN is only $\simeq$10$^{-10}$ \cmv\ in the center of this cloud, corresponding to a fractional abundance relative to H$_2$ of 3$\times$10$^{-13}$.  $X$(HCN) falls rapidly below this value for  $A_{\rm v}$ $\le$ 0.2 mag.  

For C$^+$ itself, the situation is quite different.  The deexcitation rate coefficients for collisions with electrons \citep{Wilson02} are $\simeq$ 350 times larger than those for collisions with atomic hydrogen \citep{Barinovs05}, and $\simeq$ 100 times greater than those for collisions with H$_2$ with an ortho to para ratio of unity \citep{Wiesenfeld14}.  Thus, even with atomic carbon totally ionized, collisions with electrons will be unimportant compared to those with hydrogen, whether in atomic or molecular form.  In fully ionized gas, on the other hand, excitation of C$^+$ will be via collisions with electrons.
%%%%%%%%%%%%%%%%%%%%%%%%%%%%%%%%%%%%%%%%%%%
%%%%  FIGURE 6
%%%%%%%%%%%%%
%%%%%%%%%%%%%%%%%%%%%%%%%%%%%%%%%%%%%%%%%%%%%%

\subsection{Electron Density in GMC Cloud Edges}
\label{edges} 
Giant molecular clouds (GMCs) exist in a large range of masses, sizes, and radiation environments, making it difficult to draw specific conclusions about the electron density within them, which varies significantly as function of position. We are interested primarily in the outer portion of the cloud where we expect the electron fractional abundance to be maximum.  Such regions are, in fact the Photon Dominated Region (PDR) that borders every such cloud.  As an illustrative example, we consider a slab cloud with a thickness equal to 5$\times$ 10$^{18}$ cm and a Gaussian density distribution with central proton density equal to 1$\times$10$^5$ \cmv\ and 1/e radius  2.35$\times$10$^{18}$ cm, leading to an edge density equal to 1.1$\times$10$^3$ \cmv.  The total cloud column density measured normal to the surface is 4.2$\times$10$^{23}$ \cmc.  

\begin{figure}[h!]
\begin{center}
\includegraphics[angle=0, width=\linewidth]{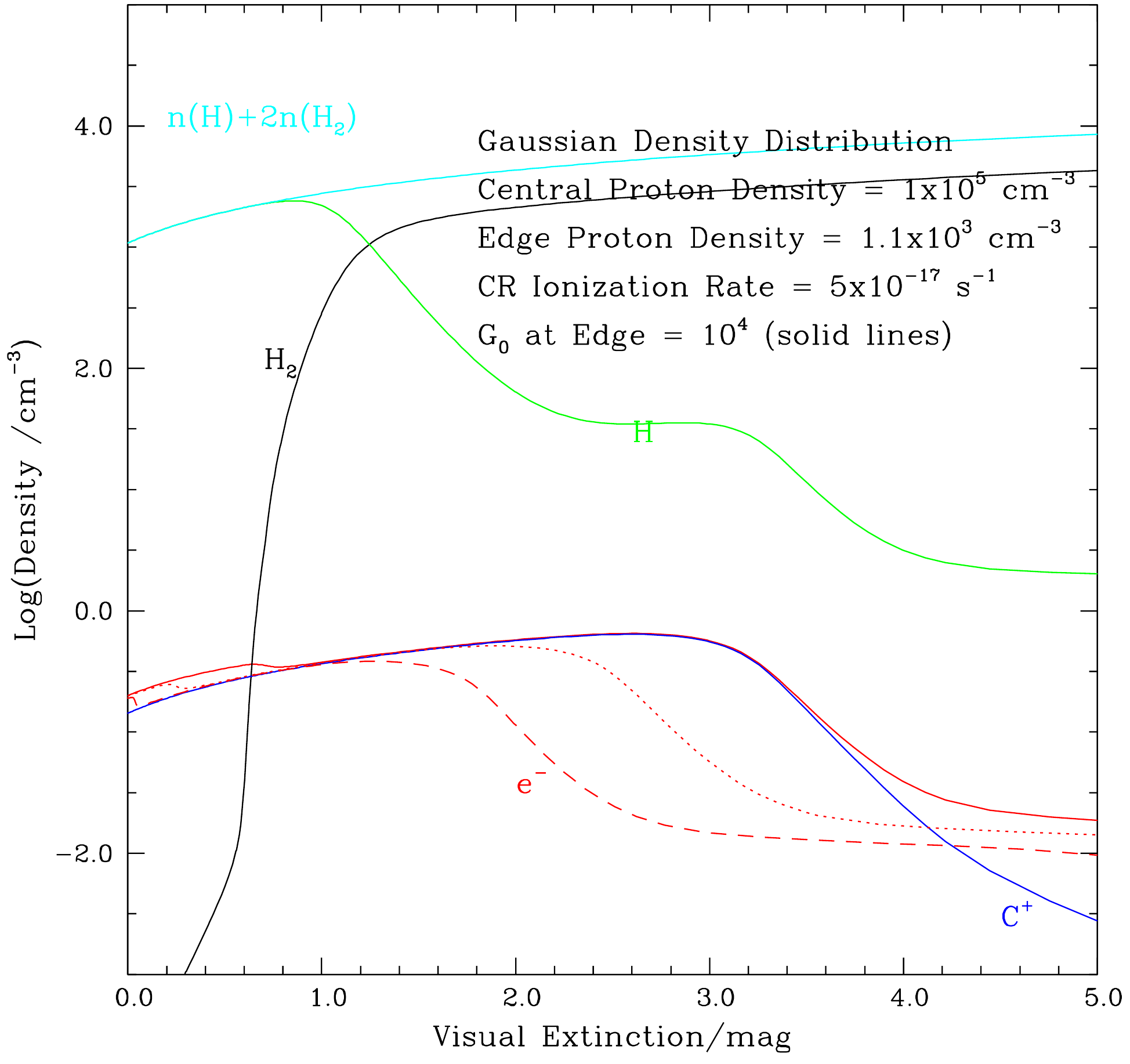}
\caption{Edge region of a cloud with Gaussian density distribution having parameters given in the Figure.   The solid curves are for radiation field enhanced by a factor of 10$^4$ relative to standard ISRF.   Ionized carbon is the dominant source of electrons, and this region encompasses the outer 3 magnitudes of the cloud; $X$(e$^-$) $\simeq$ 2$\times$10$^{-4}$ for A$_{\rm v}$ $\leq$ 2.6 mag., and drops to $\simeq$ 10$^{-5}$ at A$_{\rm v}$ = 4 mag.  The dotted and dashed curves describe the electron density for radiation field enhancements relative to the standard ISRF of factors 10$^3$ and 10$^2$, respectively, and show that the region of large $n$(e$^-$) becomes more limited as the incident radiation field intensity is reduced.}
\label{gausscloud_comb}
\end{center}
\end{figure}

The results from the Meudon PDR code are shown in Figure \ref{gausscloud_comb}. The solid curves are for a radiation field a factor of 10$^4$ greater than the standard ISRF.   In the outer portion of the cloud shown, the transition from H to H$_2$ occurs at a density of $\sim$ 10$^3$ \cmv\ and an extinction of $\simeq$ 1.2 mag, as a result of the relatively high external radiation field incident on the surface of the cloud.  This level of radiation field is not unreasonably large for a cloud in the vicinity of a massive young stars.  Using the model of \citet{Stacey93}, the front surface of the Orion cloud within a radius of $\sim$0.9 pc of the Trapezium cluster is subject to a radiation field of this or greater intensity.  

In the outer portion of the cloud, the electron density is essentially equal to that of C$^+$, and the fractional abundance $X({\rm e}^-)$  $\simeq$ 2$\times$10$^{-4}$ in the outermost 1.2 mag., where atomic hydrogen is dominant, and remains at this value to the point where $A_{\rm v}$ = 3 mag.   The electron density drops significantly moving inward from this point, falling to 10$^{-5}$ at $A_{\rm v}$ = 4 mag.  From Figure \ref{elec_1e31e4}, we see that the electrons increase the emission in HCN $J$= 1--0 by an order of magnitude relative that from H$_2$ at the point where $n$(H$_2$) $\simeq$10$^3$ \cmv.  

The dotted and dashed curves show the electron density for radiation fields increased by factors 10$^3$ and 10$^2$, respectively, relative to the standard ISRF.  The lower radiation fields reduce the thickness of the layer of high electron density, but only slightly affect the density there.  Reduction in the radiation field by a factor of 100 reduces the thickness of the layer by $\simeq$ a factor of 2 in terms of extinction.  

The results from this modeling suggest that regions of significant size can have electron densities sufficient to increase the excitation of any high--dipole moment molecules that may be present by a factor $\simeq$ 5.  An important requirement for this to be of observational significance is that the density of the species in question be sufficient in the region of enhanced electron density.  This issue is discussed in the following section.

%%%%%%%%%%%%%%%%%%%%%%%%%%%%%%%%%%%%%
%%%%   FIGURE 7
%%%%%%%%%%%%%%

%%%%%%%%%%%%%%%%%%%%%%%%%%%%%%%%%%%%%%%

\subsection{Molecular Abundances in GMC Cloud Edges}
\label{chem}
\begin{figure}[h]
\begin{center}
\includegraphics[angle=0, width=\linewidth]{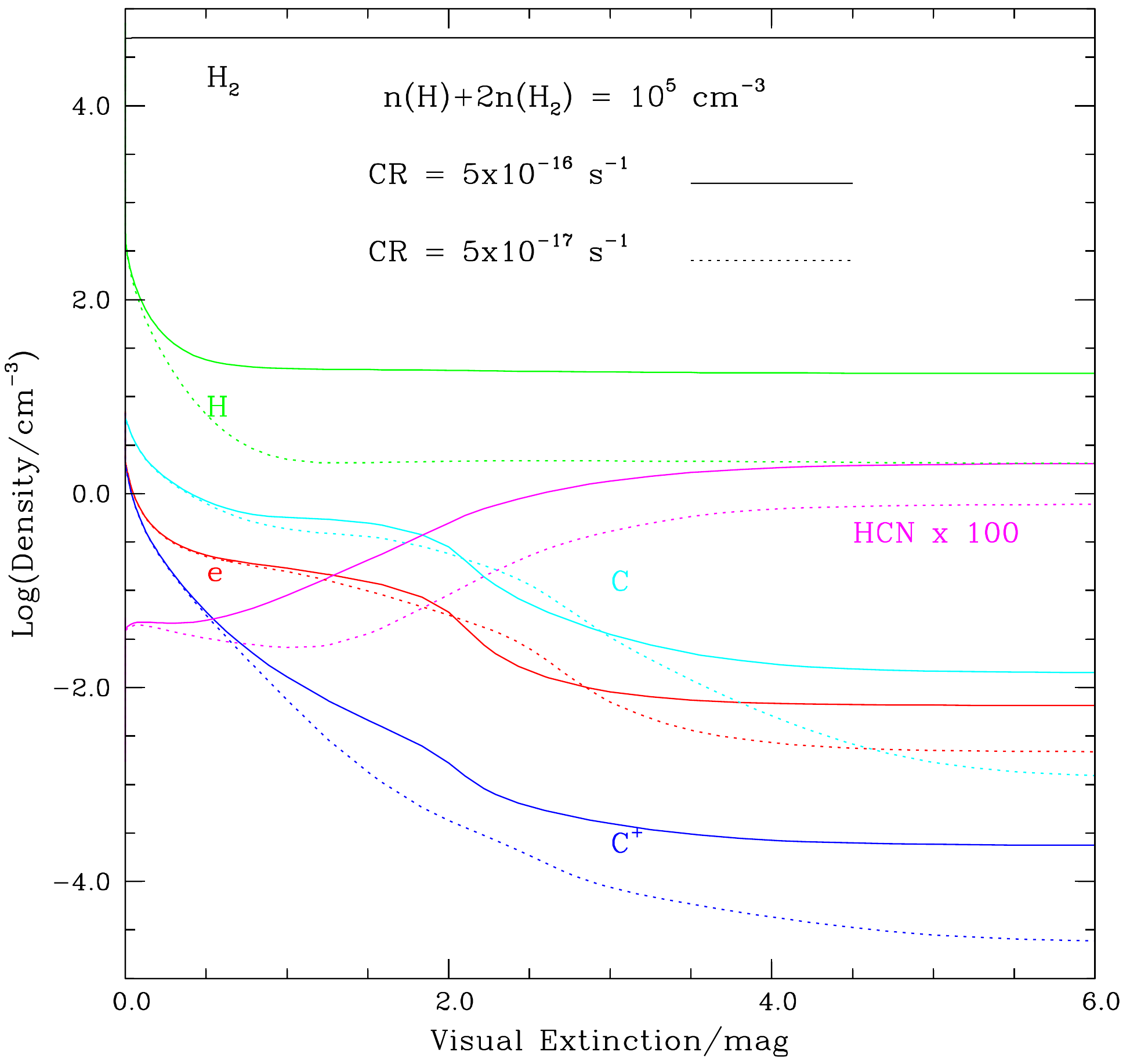}
\caption{Effect of cosmic ray ionization rate on the outer portion of a cloud having uniform proton density = 10$^5$ \cmv\ and total extinction = 50 mag.  The incident radiation field is the standard ISRF.  The dotted curves are for a standard cosmic ray ionization rate of 5$\times$10$^{-17}$ s$^{-1}$, while the solid curves are for a rate a factor of 10 higher.  The higher cosmic ray ionization rate has very little effect on the electron density in the outer few mag, where it is very close to the abundance of C$^+$.  An enhanced cosmic ray ionization rate does increase the HCN abundance by a factor of a few for $A_{\rm v}$ $\ge$ 1 mag. 
}
\label{cloudedge}
\end{center}
\end{figure}

Standard (e.g. the Meudon PDR code utilized in \S \ref{edges}) models of the chemistry in low--extinction portions of interstellar clouds predict that the density and fractional abundance of HCN will be quite low, as illustrated in Figure \ref{diffuse} discussed in \S \ref{diffuse_trans}.   A result for a more extended, higher density region, also obtained using the Meudon PDR code, is shown in Figure \ref{cloudedge}, which focuses on the outer region of a cloud with uniform proton density = 10$^5$ \cmv\ and total extinction = 50 mag.  The incident radiation field is the standard ISRF.  Within 2 mag of the cloud boundary we find $X$(HCN) $\sim$ 4$\times$10$^{-9}$, a factor $\simeq$ 40 less than that in the region with $A_{\rm v}$ $\ge$\ 4 mag.  This is sufficiently small to make the emission per unit area from the outer portion of the cloud, even with electron excitation, relatively weak relative to that in the better--shielded portion of the cloud.  However, depending on the structure of the cloud and the geometry of any nearby sources enhancing the external radiation field, the electron excitation could significantly increase the total high--dipole moment molecular emission from the cloud.  As indicated in Figure \ref{cloudedge}, an increased cosmic ray ionization rate does increase the HCN abundance in portions of the cloud characterized by $A_{\rm v}$ $\ge$ 1 mag.  The larger rate here is a reasonable upper limit for the Milky Way, so this effect is likely to be limited, but not necessarily in other galaxies \citep[e.g.][]{Bisbas15}.

The abundance of high--dipole moment molecules (e.g. HCN and CS) and that of electrons, as enhanced by either UV or cosmic rays, are to a significant degree anticorrelated in standard PDR chemistry.  This is illustrated in Table \ref{anti} which gives some results for PDR models of clouds of different densities with total visual extinction equal to 50 mag, illuminated from both sides by standard ISRF, and experiencing a cosmic ray ionization rate equal to 5$\times$10$^{-16}$ s$^{-1}$.  This enhanced rate is responsible for the relatively large densities of atomic hydrogen in the well--shielded portions of the cloud.

%%%%%%%%%%%%%%%%%%%%%%%%%%%%%%%%%%%%%
%%%%   TABLE 3
%%%%%%%%%%%%%%%%%%%%
\begin{deluxetable*}{cccc}
\tablewidth{0pt}
\tablecaption{Densities of Different Species in Clouds of Different Densities \label{anti}
} 
\tablehead{  
\colhead{Visual Extinction} & {$n$(H)} &{$n$(e$^-$)} & {$n$(HCN)}\\
\colhead{mag}  & {cm$^{-3}$} & {cm$^{-3}$} & {cm$^{-3}$}
 }
\startdata
\multicolumn{4}{c}{n(H) + 2n(H$_2$) = 10$^6$ cm$^{-3}$}\\[0.5ex]
0.1 & 5.1$\times$10$^3$ & 2.0$\times$10$^{+0}$ & 1.2$\times$10$^{-2}$\\
1.0 & 1.9$\times$10$^1$ & 7.4$\times$10$^{-1}$ & 9.2$\times$10$^{-3}$\\
3.0 & 2.1$\times$10$^1$ & 2.5$\times$10$^{-2}$ & 6.4$\times$10$^{-2}$\\[0.5ex]
\tableline\\[-1.5ex]
\multicolumn{4}{c}{n(H) + 2n(H$_2$) = 10$^5$ cm$^{-3}$}\\[0.5ex]
0.1 & 9.6$\times$10$^1$ & 1.7$\times$10$^{-1}$ & 4.7$\times$10$^{-4}$\\
1.0 & 2.0$\times$10$^1$ & 8.0$\times$10$^{-3}$ & 9.0$\times$10$^{-4}$\\
3.0 & 1.8$\times$10$^1$ & 8.0$\times$10$^{-3}$ & 1.5$\times$10$^{-2}$\\[0.5ex]
\tableline\\[-1.5ex]
\multicolumn{4}{c}{n(H) + 2n(H$_2$) = 10$^4$ cm$^{-3}$}\\[0.5ex]
0.1 & 1.2$\times$10$^2$ & 6.9$\times$10$^{-1}$ & 4.2$\times$10$^{-6}$\\
1.0 & 2.1$\times$10$^1$ & 3.3$\times$10$^{-2}$ & 4.0$\times$10$^{-5}$\\
3.0 & 1.9$\times$10$^1$ & 5.5$\times$10$^{-3}$ & 1.9$\times$10$^{-3}$\\[0.5ex]
\tableline\\[-1.5ex]
\multicolumn{4}{c}{n(H) + 2n(H$_2$) = 10$^3$ cm$^{-3}$}\\[0.5ex]
0.1 & 3.7$\times$10$^{1}$ & 1.3$\times$10$^{-1}$ & 8.0$\times$10$^{-8}$\\
1.0 & 2.4$\times$10$^{1}$ & 2.9$\times$10$^{-2}$ & 8.5$\times$10$^{-7}$\\
3.0 & 1.9$\times$10$^{1}$ & 3.1$\times$10$^{-3}$ & 7.4$\times$10$^{-5}$
\enddata
\end{deluxetable*}

%%%%%%%%%%%%%%%%%%%%%%%%%%%%%%%%%%%%%
We see that only for the two lowest densities and for visual extinction less than 1 mag is the density of electrons high enough to significantly increase the excitation rate.  However, the HCN density under these conditions corresponds to a fractional abundance only 1/1000 of that which can be reached in the well--shielded portions of clouds.  Thus, the effect of the electron enhancement of the collision rate would be very difficult to discern.  At high hydrogen densities, the density of electrons increases but their abundance relative to H$_2$, the dominant form of hydrogen is quite low.  Thus, even though the HCN density increases almost as the square of the total density at low extinctions in this model, the HCN emission when strong, will be produced by collisions with H$_2$. 
%%%%%%%%%%%%%%%%%%%%%%%%%%%%%%%%%%%%%
%%%%%%%   FIGURE 8
%%%%%%%%%%%%%%%%%%%%
%%%%%%%%%%%%%%%%%%%%%%%%%%%%%%%%%%%%%%%

The issue of the  fractional abundance of all molecules in all regions may not yet be treated completely by such models.  For example, the abundance of CO in diffuse clouds is well known to be much greater relative to standard models, and a variety of processes involving transient high temperatures have been proposed \citep{Elitzur78, Federman96, Zsargo03}; see also Section 4.4 of \citet{Goldsmith13}.  Possible mechanisms responsible for the elevated temperature include shocks, Alfv\'{e}n waves, and turbulent dissipation \citep{Godard09}.

The situation in GMCs is even less clear as their range of densities and other physical conditions makes determination of abundance of a specific species at a particular position in a cloud very difficult.  However, if any or all of the above processes suggested to operate in diffuse clouds also are present in the outer regions (or possibly the entire volume) of GMCs, molecular abundances may also be significantly different than would expected from models with chemistry determined by the local kinetic temperature.  The possibility of significant additional energy input to the gas in the boundary of the Taurus molecular cloud, a region with relatively low radiative flux, is suggested by the detection of emission in the rotational transitions of H$_2$, indicating that temperatures of several hundred K are present \citep{Goldsmith10}.  

 Questions such as the apparent high abundance of atomic carbon throughout the volume of clouds \citep{Plume00, Howe00}, in contrast to what is predicted by chemical models with smoothly--varying density \citep[e.g.][]{Tielens85}, has motivated creation of highly--inhomogeneous ``clumpy'' cloud models \citep{Meixner93, Stoerzer96, Roellig06, Cubick08}.  In this picture, UV photons can permeate a large fraction of the clouds' volume, producing PDR regions on the clumps distributed throughout the cloud.  Thus, the regions of high electron density which are located adjacent to where the C$^+$ transitions to C, are also distributed throughout the cloud.

An entirely different class of models involves large--scale circulation of condensations between the outer regions of clouds and their interiors \citep{Boland82, Chieze89}.  The effect of the circulation depends on many parameters, in particular the characteristic timescale, which is not very well determined.  Yet another effect that may be significant is turbulent diffusion, which can significantly affect the radial distribution of abundances if the diffusion coefficient is sufficiently rapid \citep{Xie95}.  HCN and electrons are not included specifically in the results these authors present, but given the nature of the mechanism, it is likely that the distribution of these species, with the former centrally concentrated and the latter greater at the edge in the absence of turbulent diffusion, will be made more uniform.

If the abundance of HCN (and other high density tracers) is reasonably large in the outer portion of molecular clouds, and the electron fractional abundance approaches or exceeds 10$^{-4}$ there, the total mass of the ``high density region'' may be overestimated.  This could have an impact on using such molecular transitions as tracers of the gas available for the formation of new stars \citep[e.g.][]{Gao04}, and the possible role of electron excitation in enhancing the size of the high--dipole moment molecular emission should be considered.

\subsection{Extreme Cloud Environments}
\label{extreme} 
The central regions of both starburst and AGN galaxies are extreme environments, with dramatically enhanced energy inputs compared to the ``normal'' ISM of the Milky Way and normal galactic disks.  Determining the conditions in these regions is naturally challenging, but with the increasing availability of interferometers such as ALMA, there has been heightened interest in unravelling the physical conditions in the central nuclear concentration(s) as well as the surrounding tori that are seen in some galaxies.  The density is one of the most important parameters, and low--$J$ transitions of HCN and its intensity relative to other species, are one of the most often--used probes.  

The ratio of HCN to CO and HCN to HCO$^+$  in the lowest rotational transition of each were observed in a number of nearby Seyfert galaxies by  \citet{Kohno01}, who proposed that the observed enhancement of $R$ = $I$(HCN)/$I$(CO) in those dominated by an AGN could be produced by enhanced X--ray irradiation of the central region, based on the modeling of \citet{Lepp96}.  This interpretation was used to explain observations of the AGN NGC1068 by \citet{Usero04}.  This diagnostic was extended to the $J$=3--2 transitions of HCN and HCO$^+$ in NGC1097 by \citet{Hsieh12}, who found that the enhanced ratio of these two molecules was consistent with X--ray ionization, using the model of \citet{Maloney96}.  More detailed models of X--ray dominated regions have since been developed by \citet{Meijerink05} and \citet{Meijerink07}, which indicate a different effect on the HCN/HCO$^+$ ratio than found by \cite{Maloney96}.

However, the X--ray ionization and heating is not the only mechanism proposed to explain enhanced HCN emission.  Heating alone, if sufficient to accelerate the reaction CN + H$_2$ $\rightarrow$ HCN + H ($\Delta E/k$ = 820 K), can increase the abundance of HCN.  \citet{Aalto12} proposed that shocks could be compressing and heating the gas in the outflow associated with AGN Mrk 231.  \citet{Izumi13} observed the $J$=4--3 transition of HCN and HCO$^+$, along with other molecules, in the nucleus of AGN NGC1097. Combining their data with others (their Table 7) indicates that the HCN enhancement is greater in AGN than in Starburst galaxies.  Their chemical modeling suggests a dramatically enhanced HCN abundance based solely on having the gas temperature exceed 500 K.  These authors reject UV and X--rays as the explanation and favor mechanical heating, possibly from a (so far unobserved) AGN jet.  \citet{Martin15} also favor non--X--ray heating to explain their observations of the galaxy Arp 220, employing the chemical models of \citet{Harada10} and \citet{Harada13}.

In the context of these observations and proposed models, the relevance of electron excitation of high dipole moment molecules such as HCN is that the regions of the enhanced HCN abundance, whether produced by X--rays, shocks, or UV, could well include substantial electron densities as well.  We discussed previously that the integrated intensities of HCN emission could be substantially enhanced if the fractional abundance of electrons is on the order of 10$^{-4}$.  

For subthermal excitation and optically thin emission, the $J$ = 1--0 integrated intensity (equation \ref{Tdv_crit}) can be written
\beq
\label{extreme_1}
\int T_{\rm a} dv \propto N({\rm molecule}) A_{10} \frac{n({\rm H}_2)}{n_c({\rm H}_2)} \left( 1 + \frac{X({\rm e}^-)}{X^*({\rm e}^-)} \right) \lp
\eeq
If we consider a given molecular species in a region of specified H$_2$ density, the effect of the electron excitation is contained in the second term in brackets.  As examples, for a fractional  abundance of 10$^{-5}$, the HCN and CN emission will be approximately doubled, while that of HCO$^+$ and CS will be only slightly enhanced.  For a fractional  abundance of 10$^{-4}$, the emission from HCN and CN will be increased by approximately an order of magnitude, while that of CS and HCO$^+$ by factors $\simeq$ 2.3 and 3.6, respectively.  

%For subthermal excitation and optically thin emission, the $J$ = 1--0 integrated intensity is (equations \ref{Tdv_Ct} and \ref{Ct_def}) proportional to the total excitation rate from $J$ = 0 to $J$ = 1, as well as to the column density of the species in question.  Focusing on the collision rate, we can write 
%\beq
%\label{eq:extreme-subthermal_1}
%C^{\rm t}_{01} = R^{\rm e}_{01} {\rm (H_2)n(H_2)}
%\left[1 + \frac{ R^{\rm e}_{01}(\rm e{^-})}   {R^{\rm e}_{01}({\rm H}_2)}
%	\frac{{\rm n(e^-})} {{\rm n(H_2)}} \right] \lc
%\eeq
%or
%\beq
%\label{eq:extreme-subthermal_2}
%C^{\rm t}_{01} = {\rm H_2~Excitation~Rate~} 
%\left[1 + \frac{ R^{\rm e}_{01}(\rm e{^-})}   {R^{\rm e}_{01}({\rm H}_2)}
%	X({\rm e^-})  \right] \lp
%\eeq

%HCO$^+$ has rate coefficients for collisions with H$_2$ that are a factor $\simeq$ 20 larger than those for HCN \citep{Flower99}, while as discussed in \S\ref{HCO+dat}, the rate coefficients for collisions with electrons are a factor $\simeq$3 larger.  The result is that the term multiplying the electron fractional abundance is approximately a factor of 7 times smaller for HCO$^+$ than for HCN.  Thus, for an electron fractional abuance of 10$^{-4}$, the HCN excitation rate (and thus emission) would be increased by a factor of 10 compared to that produced by the local H$_2$ density, while that for HCO$^+$ would be increased only by a factor of 2--3.  

Electron excitation could thus be responsible at least in part for the enhanced HCN/HCO$^+$ ratio reported by \citet{Kohno01} in some Seyfert nucleii.  For CO (\S \ref{rates}), the electron excitation rates are dramatically smaller than those for HCN so that X(e$^-$) $\geq$ 10$^{-5}$  will dramatically enhance HCN emission relative to that of CO, which \citet{Kohno01} report to be correlated with enhanced HCN/HCO$^+$.

\begin{figure}[h!]
%\begin{center}
\includegraphics[angle=0, width=\linewidth]{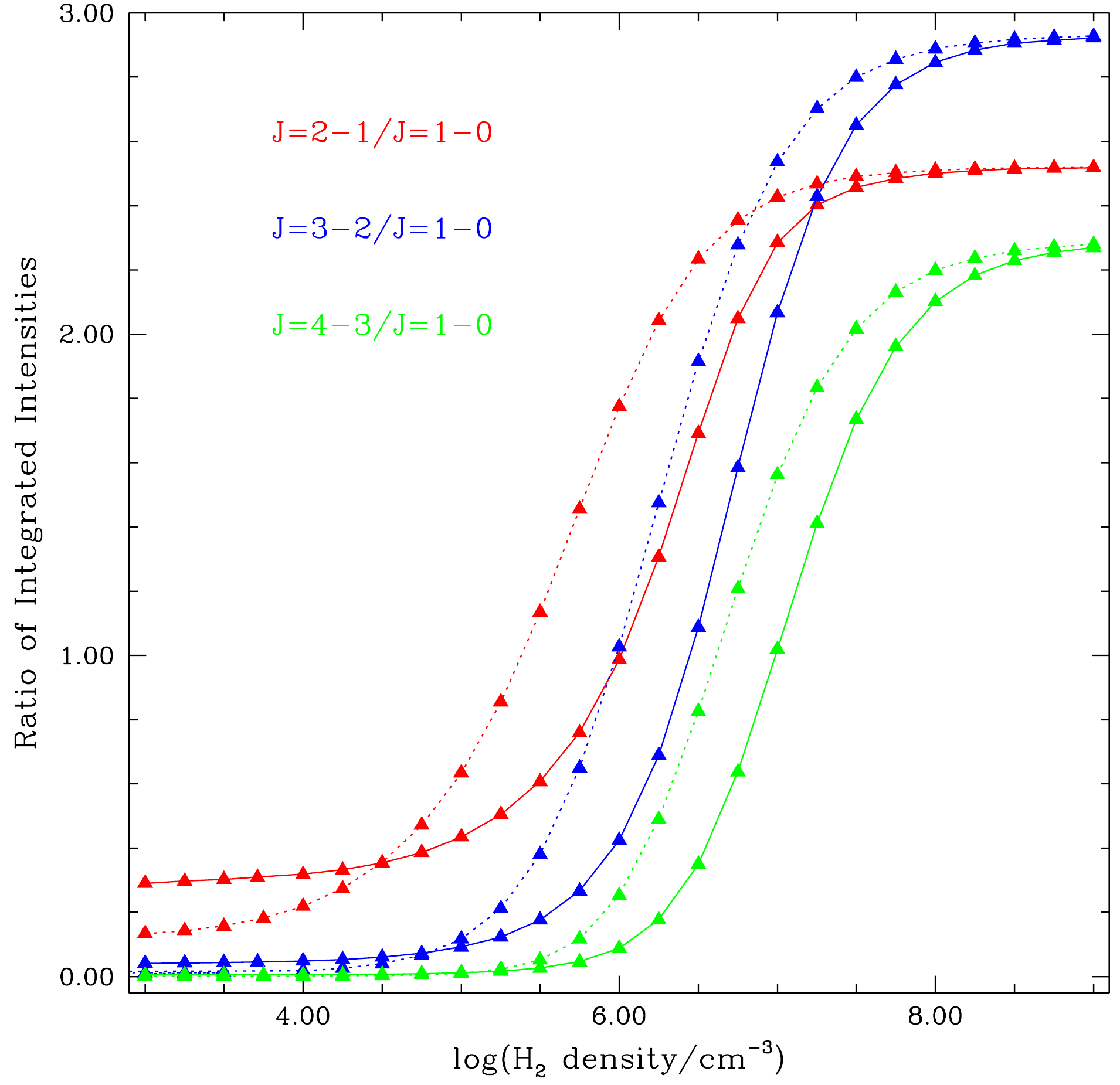}
\caption{Ratio of integrated antenna temperatures of higher rotational transitions of HCN to that of the $J$=1-0 transition.  The different colors are for the three higher transitions indicated.  The kinetic temperature is 20 K, the column density of HCN is 1$\times$10$^{13}$ \cmc, and the line width is 2.5 \kms.  The solid lines give the ratio when collisions are exclusively with H$_2$.  The dotted curves give the ratio when the H$_2$ is accompanied by a fractional abundance of electrons $X$(e$^-$) = 1$\times$10$^{-4}$.  The presence of the electrons reduces the H$_2$ density required to achieve a moderate ratio, 0.5 -- 2.0, by a factor of 2 to 4 depending on the higher transition in question. }
\label{ratio}
%\end{center}
\end{figure}

\citet{Izumi13} employed the ratio of different HCN transitions to determine volume densities of H$_2$ and properties of the HCN--emitting region.  Electron excitation can produce different ratios than does H$_2$ excitation as a result of the different $J$--dependence of the collision rate coefficients (Appendix~\ref{app:simple} and \citet{Dickinson77}). Figure \ref{ratio} shows the ratio of integrated intensities of different transitions $J$ -- $J$--1 relative to that of the 1--0 transition.  A 10--level calculation using RADEX was employed.  The presence of a fractional abundance of electrons equal to 10$^{-4}$ reduces the H$_2$ density to achieve a specified integrated intensity ratio.  Observed ratios yield H$_2$ densities in the range 10$^5$ to 10$^6$ \cmv.  $X$(e$^-$) = 10$^{-4}$ reduces the required H$_2$ density by a factor 3 to 4, which would have a significant impact on characterizing the central regions of galaxies.  How important and prevalent the effect of electron excitation is depends on having more reliable models of the radiation field, ionization and chemistry in the central regions of these luminous galaxies.  Many effects that may be playing a role.  IR pumping is likely important in some sources, as indicated by the detection of vibrationally excited HCN by \citet{Aalto15a, Aalto15b}. \citet{Bisbas15} studied the effect of cosmic ray ionization rates up to a factor of 10$^3$ greater than standard (as compared to the modest factor of 10 considered in Figure \ref{cloudedge}) on the CO/H$_2$ ratio.  The CO/H$_2$ ratio is reduced, but the magnitude of the effect depends on the local density.  This study suggests that turbulent mixing, although not included in the modeling, is potentially important.

%%%%%%%%%%%%%%%%%%%%%%%%%%%%%%%%%%%%%%%%%%%
%%%%   FIGURE 9
%%%%%%%%%%%%%%%%%%%
%%%%%%%%%%%%%%%%%%%%%%%%%%%%%%%%%%%%%%%

\section{Conclusions}
\label{conc}
We have used quantum calculations of collisional excitation of the rotational levels of HCN, HCO$^+$, CN, and CS by electrons and H$_2$ molecules to evaluate the relative importance of electron excitation. The collisional deexcitation rate coefficients at the temperatures of molecular clouds are close to 10$^5$ times larger for electrons than for H$_2$ molecules (\S{}\ref{sec:critical-fractional-abundance}).  The electron deexcitation rate coefficients scale as the square of the permanent electric dipole moment of the target molecule, so this effect is unimportant for the widely--used tracer CO.    For subthermal excitation, the integrated intensity of the $J$ = 1--0 transition is proportional to the sum of the electron and H$_2$ densities each normalized to the appropriate critical density (Eq.~\ref{Tdv_crit}). 

The requirements for electron excitation to be of practical importance are $n({\rm H}_2) \leq\ n_{\rm{}c}({\rm H}_2)$ and $X({\rm e}^-) \geq\ X^*({\rm e}^-)$ (\S{}\ref{sec:emission-tex}; also see Eq.~\ref{extreme_1}). For the J = 1--0 transition of HCN this implies $n({\rm H}_2) \leq\ 10^{5.5}$ \cmv\ and $X({\rm e}^-) \geq\ 10^{-5}$. In regions where carbon is largely ionized but hydrogen is molecular, the fractional abundance of electrons, $X$(e$^-$) = $n$(e$^-$)/$n$(H$_2$) can exceed 10$^{-4}$, making electrons dominant for the excitation of HCN (\S{}\ref{edges}).  The situation for CN is similar, although somwhat more uncertain due to less complete collision rate calculations. For HCO$^+$, the rate coefficients for collisions with H$_2$ are more than an order of magnitude larger than those for HCN, more than outweighing the somewhat larger rate coefficients for electron collisions, and demanding a factor 7--10 of higher electron fractional abundance for electron excitation to be significant.  For CS, the rate coefficients for electrons are a factor of 2 smaller than those for HCN, while the H$_2$ rate coefficients are a factor $\simeq$ 3 larger, with the combination results in requirement of a factor of 6 larger electron fractional abundance for electron excitation to be significant.  Thus, HCN (and to slightly lesser degree CN) appears to be an unusually sensitive probe of electron excitation (Table~\ref{critical}).

Conditions favoring high $X$(e$^-$) can occur in low extinction regions such as diffuse and translucent clouds (\S{}\ref{diffuse_trans}), and the outer parts of almost any molecular cloud, especially in regions with enhanced UV flux (\S{}\ref{edges}).  Thus, the excitation in the HCN--emitting region may not necessarily be controlled by the high H$_2$ density generally assumed.  The central regions of luminous galaxies often show enhanced HCN emission, which could be in part a result of electron excitation, although the explanation is not certain, with enhanced UV, X--rays, cosmic rays, and mechanical heating having all been proposed as responsible for increasing the abundance of HCN (\S{}\ref{extreme}).

Accurate determination of the possible importance of electron excitation will depend on having much improved models of the chemistry and dynamics of these regions, including the effects of transient heating and enhanced transport due to turbulence (\S{}\ref{chem}). Significant additional theoretical work is therefore needed before a satisfying explanation can be given for the extended emission from molecules like HCN in low density environments \citep{Kauffmann17}.

%\acknowledgments %ApJ style%%%%%%%%%%%%%%%%%%%%%%%%%%%%%%%%%%%%%%
\begin{acknowledgements}

We thank Simon Glover for suggesting that we consider electron excitation in the outer parts of molecular clouds.  We thank the anonymous reviewer for suggestions that significantly broadened and improved the present investigation. We appreciate Floris van der Tak's critical help in entering data into and using RADEX, and thank Franck LePetit for valuable information concerning the Meudon PDR code.   The authors appreciate information and pointers received  from Susanne Aalto about molecules in Active Galactic Nucleii.  Alexandre Faure graciously provided the full unpublished HCO$^+$--electron deexcitation rate coefficients from unpublished work by Faure and Tennyson.  We thank Bill Langer and Kostas Tassis for a number of suggestions that improved this paper.   This research was conducted in part at the Jet Propulsion Laboratory, which is operated by the California Institute of Technology under contract with the National Aeronautics and Space Administration (NASA). \copyright2016 California Institute of Technology. \\

\end{acknowledgements}

%%%%%%%%%%%%%%%%%%%%%%%%%%%%%%%%%
\bibliography{./bibdata}

\appendix

\section{Simplified Models of Excitation}
\label{app:simple}

In this Appendix we outline a very simplified model with which to give an idea of the relative importance of collisional excitation by electrons and by molecular hydrogen in the limit of low collision rates.  We again adopt HCN as a representative high--dipole moment molecule.  We adopt the rate coefficients for collisions with H$_2$ from \citet{Dumouchel10}, understanding that while the result of \citet{BenAbdallah12} are quite similar, the situation could be quite different if a large fraction of the H$_2$ were in states having $j$ $\ge$ 1 and the results of \citet{Vera14} discussed above obtain.

It is instructive to consider only a three level system (levels and rotational quantum numbers $J$ = 0, 1, and 2) with no background radiation and optically thin transitions. 
The collision rates $C_{ij}$ (s$^{-1}$) are equal to the collisional rate coefficients $R_{ij}$ (cm$^3$s$^{-1}$) multiplied by the density of colliding particles (electrons or H$_2$ molecules, \cmv).  In general we must consider upwards and downwards collisions, but in the limit of low excitation, with the spontaneous downwards radiative rate, $A_{ul}$, much larger than the corresponding downwards collision rate, $C_{ul}$, downwards collisions can be neglected.  The rate equations for the densities of HCN, $n_0$, $n_1$ and $n_2$ in  levels $J$= 0, 1 and 2, respectively, are
\beq
\label{rate1}
n_0(C_{01} + C_{02}) = n_1A_{10} \lc
\eeq

\beq
n_1A_{10} = n_0C_{01} + n_2A_{21} \lc
\eeq
and
\beq
n_2A_{21} = n_0C_{02} + n_1C_{12}   \lc
\eeq
where $C_{lu}$ denotes the upward collision rate from level $l$ to level $u$.    Equation \ref{rate1} gives us immediately the ratio 
\beq
\label{lowdenratio}
\frac{n_1}{n_0} = \frac{C_{01} + C_{02}}{A_{10}} \lp
\eeq
For more than three levels in the low excitation limit, it is appropriate to consider the total upwards collision rate out of $J$ = 0 when analyzing the excitation of the $J$ = 1 to $J$ = 0 transition, as every such collisional excitation results in emission of a $J$ = 1 to $J$ = 0 photon.  We define an effective excitation rate 
\beq
C^e_{01}({\rm H}_2) = C_{01}({\rm H}_2) + C_{02}({\rm H}_2) \lc
\eeq
for 3 levels, and 
\beq
\label{geneffective}
C^e_{01}({\rm H}_2) = \Sigma_{k = 1}^{k =N}C_{0k}({\rm H}_2)
\eeq
for $N$ levels, since collisions with H$_2$ can result in $|\Delta J|$ $>$ 1.   We can express this as well in terms of effective rate coefficients since we have only to divide by the density of collision partners, and for the deexcitation rate coefficients we have $R^e_{10}({\rm H}_2) = C^e_{10}({\rm H}_2)/n({\rm H}_2)$.  From detailed balance for collisions with any partner
\beq
R^e_{10} = R^e_{01}\frac{g_0}{g_1} \exp(T^*_{10}/T_k) \lc
\eeq
where the $g's$ are the statistical weights, $T^*_{10}$ is the equivalent temperature of the $J$ = 1 to $J$ = 0 transition ($\Delta E/k_B$ = 4.25 K for HCN), and $T_k$ is the kinetic temperature.  Published calculations generally give the downwards rate coefficients, and the upwards rate coefficients must be calculated individually using detailed balance.  

 For collisions with H$_2$  at a kinetic temperature of 20 K, \citet{Dumouchel10} give $R_{10}$ = 1.41$\times$10$^{-11}$  cm$^3$s$^{-1}$ and $R_{20}$ = 2.1$\times$10$^{-11}$ cm$^3$s$^{-1}$.  $R_{30}$ is 100 times smaller than these rate coefficients, while $R_{40}$ = 2.7$\times$10$^{-12}$ cm$^3$s$^{-1}$ is marginally significant.  From each of these we calculate the upwards rate coefficient, and the effective downward rate coefficient is 
$R^e_{10}$(H$_2)$ = 3.8$\times$10$^{-11}$ cm$^3$s$^{-1}$ and
$R^e_{01}$(H$_2)$ = 9.2$\times$10$^{-11}$ cm$^3$s$^{-1}$.

For electrons, since we consider only dipole--like collisions, we have 
\beq
C^e_{10}({\rm e}^-) = C_{10}({\rm e}^-) = R_{10}({\rm e}^-)n({\rm e}^-) \lc
\eeq
 where  $R_{10}({\rm e}^-)$ is just the value from Table 1 at the appropriate kinetic temperature, which is 3.7$\times$10$^{-6}$ cm$^3$s$^{-1}$ at 20 K. 

The critical density $n_c$ is the density of colliding partners at which the downwards collision rate is equal to the spontaneous decay rate.  This gives us for the $J$ = 1--0 transition of HCN
\beq
\label{elec_nc}
n_{\rm c}({\rm e}^-) = \frac{A_{10}}{R^e_{10}({\rm e}^-)} = 6.5~ {\rm cm}^{-3}\lc
\eeq
and
\beq
\label{H2_nc}
n_{\rm c}({\rm H}_2) = \frac{A_{10}}{R^e_{10}({\rm H}_2)} = 6.5\times 10^5~{\rm cm}^{-3}\lp
\eeq
Table \ref{critical} gives values for other molecules.

\end{document}